\documentclass[%
aip,
jcp,
reprint,
 amsmath,amssymb,
]{revtex4-1}

\usepackage{graphicx}
\usepackage{dcolumn}
\usepackage{bm}
\usepackage{bbold}

\begin{document}
\author{Jakub K. Sowa}
\email{jakub.sowa@northwestern.edu}
\author{Emily A. Weiss}
\author{Tamar Seideman}
\email{t-seideman@northwestern.edu}
\affiliation
{Department of Chemistry, Northwestern University, Evanston, IL 60208, USA.}

\title{Photoisomerization-coupled electron transfer}

\begin{abstract}
Photochromic molecular structures constitute a unique platform for constructing molecular switches, sensors and memory devices. One of their most promising applications is as light-switchable electron acceptor or donor units. Here, we investigate a previously unexplored process that we postulate may occur in such systems: an ultrafast electron transfer triggered by a simultaneous photoisomerization of the donor or the acceptor moiety. We propose a theoretical model for this phenomenon and, with the aid of DFT calculations,  apply it to the case of a dihydropyrene-type photochromic molecular donor. By considering the wavepacket dynamics and the photoisomerization yield, we show that the two processes involved, electron transfer and photoisomerization, are in general inseparable and need to be treated in a unified manner. We finish by discussing how the efficiency of photoisomerization-coupled electron transfer can be controlled experimentally.
\end{abstract}

\maketitle
\section{Introduction}
The phenomenon of photoisomerization, in which a molecule undergoes a structural change following its excitation with light, not only plays a critical role in the human visual perception \cite{yoshizawa1963pre,polli2010conical,mohseni2014quantum} but has also attracted attention in the field of nanotechnology where it can potentially be used to construct molecular switches, engines and optical memory devices.\cite{van2010charge,irie2000diarylethenes,yokoyama2000fulgides,kamiya2014light} 
There now exist several families of chemical compounds known to undergo reversible photoisomerization which typically involves either a ring closing/opening reaction or a \textit{cis-trans} isomerization.\cite{feringa2011molecular}
Photoisomerization of such organic compounds generally occurs through a motion of the photo-excited vibrational wavepacket through a conical intersection (CI), followed by an eventual vibrational relaxation.\cite{levine2007isomerization,polli2010conical,kazaryan2011surface,celani1997conical,domcke2004conical}

The mechanism of photoisomerization in organic molecules has been extensively studied over the past three decades and several common types of theoretical approaches can be identified. The first class of methods entails using (often high-level CASPT2/CASSCF) \textit{ab initio} calculations to construct the potential energy surface around the two minima (corresponding to the two photoisomers) and the conical intersection located between them,\cite{asano2004theoretical,ishikawa2001theoretical,quenneville2003ab} although \textit{ab initio} molecular dynamics simulations have also been used to study the non-adiabatic photoisomerization dynamics.\cite{ben1998ab,yu2015probing,yue2018performance,vreven1997ab}
Another approach, which will also be used in this work, involves modelling the   photoisomerization dynamics using idealised potential energy surfaces often also accounting for the dissipative effects arising from interactions with the broader molecular environment.\cite{hahn2000quantum,schile2019simulating,qi2017tracking,abe2005optimal,tscherbul2014excitation,thoss2006quantum}

From both a fundamental and a technological perspective, one of the most exciting applications of photochromic compounds is as light-switchable electron acceptors or donors.\cite{gust2006molecular} That is, due to the difference in oxidation (or reduction) potentials between the two isomers, photochromic compounds may act as efficient electron donors (or acceptors) only in one of their isomeric forms.
It has been suggested that such systems can form the basis of molecular optical switches and memory devices, and have therefore attracted considerable experimental interest.\cite{erno2010optical,berberich2008toward,odo2007photoswitching,berberich2012nondestructive,fukaminato2011single,fukaminato2010fluorescence,liddell2004photonic,terazono2004photonic,liddell2002photonic}

\begin{figure}[ht]
    \centering
    \includegraphics{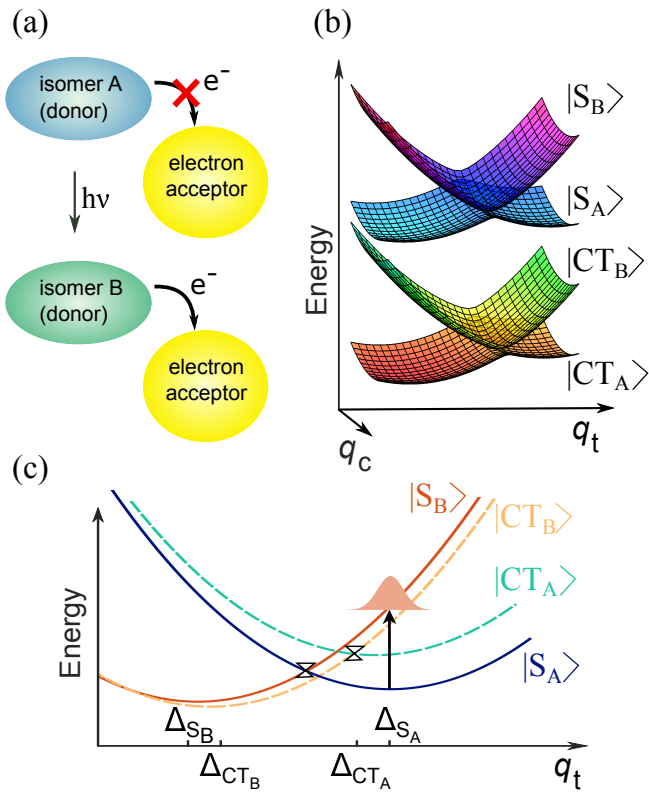}
    \caption{(a) Schematic illustration of electron transfer induced by a photoisomerization of a donor unit.  (b) Potential energy surfaces for the neutral (singlet) and charge-transfer states. For clarity, $J$ was set to zero and the charge-transfer states in (b) were shifted downwards. (c) Diabatic states plotted at $q_\mathrm{c} = 0$. CI positions are denoted above.}
    \label{fig1}
\end{figure}

Here, we suggest that photochromic electron donors (or acceptors) can also give rise to a novel phenomenon termed  \textit{photoisomerization-coupled electron transfer}.
To illustrate this concept, let us consider a photochromic unit which in one of its isomeric forms can act as an efficient electron donor, and which is connected to a (non-photochromic) acceptor unit, as schematically shown in Fig.~\ref{fig1}(a). Suppose that the photochromic donor unit was initially prepared in the isomeric form with a high oxidation potential [denoted as A in Fig.~\ref{fig1}(a)] so that the electron transfer is not possible.
One can then use a laser pulse of an appropriate wavelength to excite [and thus photoisomerise to what is denoted as isomer B in Fig.~\ref{fig1}(a)] the photochromic donor unit and hence enable the electron transfer to take place.
%
In the case of weak coupling between the donor and acceptor, the photoisomerization and the electron transfer should occur on very different timescales. First, a laser pulse excites the photochromic unit and induces its photoisomerization. Only later, the slow (non-adiabatic) electron transfer, from what is now a suitable electron donor, takes place.
In general, however, the two processes are not separable. This is especially important if the donor and acceptor units are bound covalently -- this is expected to give rise to relatively strong electronic coupling and thus a fast electron transfer, occurring on a  timescale comparable to that of photoisomerization.
This simultaneous electron transfer-photoisomerization process, in which the isomerization driving the electron transfer will itself be affected by the charge transfer dynamics, will be the focus of this work.

As stated above, photochromic electron acceptor and donor systems have been postulated as a possible basis for various optoelectronic applications such as single-molecule memory devices enabling a non-destructive readout.\cite{fukaminato2010fluorescence,berberich2012nondestructive} Such devices can, in principle, operate on ultrafast timescales and it is therefore important to understand the dynamical processes which can occur in these systems. Photoisomerization-coupled electron transfer could furthermore unlock molecular-switching functionalities not accessible in the case of `conventional' photo-switching dynamics (for which the electron transfer and photoisomerization take place sequentially).

To the best of our knowledge, neither the dynamics nor the mechanism of  photoisomerization-coupled electron transfer have been studied to date.
In this work, our primary aim is to develop a theoretical description and build an intuitive understanding of this phenomenon. 
To this end, in the latter part of this work we will consider an illustrative example of a dihydropyrene-type photochromic electron donor.

\section{Theoretical Model}
We assemble a minimal (but therefore also a general) description of the problem at hand.
Our system comprises a photochromic donor (acceptor) unit electronically coupled to an ordinary acceptor (donor) moiety.
We assume that the photochromic unit can exist in either of its photoisomeric forms, henceforth referred to as photoisomers A and B, and in either its neutral or cationic (anionic) charge state. There exist therefore four relevant electronic states: two singlet states describing neutral donor and acceptor units, $\lvert \mathrm{S_A}\rangle$ and $\lvert \mathrm{S_B}\rangle$, and two charge-transfer states (describing the cationic donor and anionic acceptor units),  $\lvert \mathrm{CT_A}\rangle$ and $\lvert \mathrm{CT_B}\rangle$.
We assume that (de-)charging of the non-photochromic unit is not accompanied by a significant displacement of its nuclear coordinates. This should be a reasonable assumption especially if that moiety corresponds to a semiconductor quantum dot.\cite{erno2010optical}

As stated before, the photoisomerization of the photochromic unit involves a motion of a photoexcited wavepacket through a CI. The minimal model for the neutral photochromic moiety comprises therefore two diabatic singlet states (corresponding to isomers A and B) coupled to the so-called tuning and coupling vibrational modes.\cite{kouppel1984multimode} The Hamiltonian for the neutral states is therefore  ($\hbar=1$):\cite{domcke2004conical,kouppel1984multimode,schile2019simulating}
\begin{equation}
    H_{0} = \sum_{i,i' = \mathrm{S_A, S_B}} \lvert i \rangle h_i \delta_{i,i'} + {\lambda_0} \: q_\mathrm{c} (1-\delta_{i,i'}) \langle i'\rvert ~,
\end{equation}
where $h_i = \varepsilon_{i} + {\kappa_\mathrm{t}^{(i)}}\: q_\mathrm{t} + T$.
In the above, $q_\mathrm{t}$ and $q_\mathrm{c}$ are the coordinates of the tuning and coupling mode, respectively,  $\varepsilon_i$ is the energy of state $i$ (at the tuning vibrational coordinate $q_\mathrm{t}=0$), ${\kappa_\mathrm{t}^{(i)}}$ is the coupling constant of the $i$-state to the tuning mode, and $\lambda_0$ determines the coupling between the diabatic states $\mathrm{S_A}$ and $\mathrm{S_B}$. The minimum of each of the diabatic states is given by $\bar{\varepsilon}_i = \varepsilon_i - \left({\kappa_\mathrm{t}^{(i)}}/\sqrt{2\omega_\mathrm{t}} \right)^2$ with a corresponding dimensionless displacement: $\Delta_i = -{\kappa_\mathrm{t}^{(i)}}/\omega_\mathrm{t}$. 
Finally, $T$ is the total vibrational operator:
\begin{equation}
    T = \sum_{k=\mathrm{c},\mathrm{t}} \dfrac{\omega_k}{2} \left( - \dfrac{\partial^2}{\partial q_k^2} + q_k^2\right)~, 
\end{equation}
where $\omega_k$ is the frequency of the mode $k$. We assumed that both the tuning and the coupling mode can be described using harmonic potentials. In reality, the tuning mode (which acts here as the reaction coordinate) may be expected to exhibit a large degree of anharmonicity. Nonetheless, we expect the model presented here to capture the essential physics of the considered phenomenon.


A similar potential energy surface (including a conical intersection) can  also be constructed for open-shell photochromic systems.\cite{sumita2006ab,mendive2016accidental} The Hamiltonian for the charge-transfer states (with a charged photochromic unit) can therefore similarly be written as:
\begin{equation}
    H_{1} = \sum_{j,j' = \mathrm{CT_A, CT_B}} \lvert j \rangle h_j \delta_{j ,j'} + {\lambda_1} \: q_\mathrm{c} (1-\delta_{j,j'}) \langle j'\rvert~,
\end{equation}
with $h_j = \varepsilon_j + {\kappa_\mathrm{t}^{(j)}}\: q_\mathrm{t} + {\kappa_\mathrm{c}^{(j)}} q_\mathrm{c} + T$.
Crucially, as it will be demonstrated below and is also shown in the supplementary material, the ground states of the charged photoisomers are in general displaced with respect to their neutral counterparts. 


The neutral and charge-transfer states are coupled electronically via
\begin{equation}
    H_{0-1} = \sum_{i,j} J_{ij} (\lvert i \rangle \langle j\rvert + \lvert j \rangle \langle i\rvert )~,
\end{equation}
where $\lvert i \rangle = \{ \lvert \mathrm{S_A} \rangle, \lvert \mathrm{S_B} \rangle \} $, $\lvert j \rangle = \{ \lvert\mathrm{CT_A} \rangle, \lvert \mathrm{CT_B} \rangle\}$, and $J_{ij}$ is the strength of the coupling between the $i$ and $j$ states.

An example of the four diabatic states considered here is plotted (as a function of the nuclear coordinates $q_\mathrm{t}$ and $q_\mathrm{c}$) in Fig.~\ref{fig1}.
The energy differences between the two photoisomers (i.e.~between $\bar{\varepsilon}_{\mathrm{S_A}}$ and $\bar{\varepsilon}_\mathrm{S_B}$, and between $\bar{\varepsilon}_\mathrm{CT_A}$ and $\bar{\varepsilon}_\mathrm{CT_B}$) are determined by the nature of the photochromic unit, see supplementary material for several examples. On the other hand, the energy differences between the neutral singlet and charge-transfer states can be tuned by choosing an appropriate (non-photochromic) donor or acceptor unit.
%

Finally, both vibrational modes are coupled to their (separate) wider environments which are modelled as collections of harmonic oscillators,\cite{kuhl2002multilevel,kuhl2000effect} see supplementary material. They account for the solvent environment of the donor-acceptor system as well as, to a lesser degree, the residual intramolecular vibrational modes.
These interactions induce vibrational relaxation (damping of the wavepacket at the rates $\gamma_k$, $k=\mathrm{c}, \mathrm{t}$) which will be modelled here using the diabatic damping approximation\cite{wolfseder1996intramolecular,may1993density,takagahara1978second} (DDA) within the Redfield theory which itself relies on the Born and Markov approximations justified in the limit of weak environmental coupling.\cite{breuer2002theory} It will be therefore assumed that the phononic environments remain thermalised at all times and can be described using continuous spectral densities. We note that  DDA has previously been used in the modelling of ultrafast electron transfer where it compared favourably with the secular approximation method.\cite{egorova2001modeling} Furthermore, unlike Redfield theory itself, DDA does not violate positivity nor does it require a detailed knowledge of the  vibrational environment.
Given the limitations of the (phenemenological) weak-coupling approach used in this work, extending the theory discussed here to include more sophisticated\cite{schile2019simulating} or even numerically exact\cite{chen2016dissipative} descriptions of environmental interactions would be an interesting, and important, avenue for future research.
We nonetheless believe that the results presented in this work grasp the fundamental physics of the considered phenomenon.
In what follows, we will also disregard any direct coupling between the electronic degrees of freedom and the outer-sphere environment which plays a lesser role in the case of non-polar solvents (as used, for instance, in Ref.~\onlinecite{erno2010optical}).

Unless specified otherwise, we will henceforth assume that the initial state was prepared by an ultrashort laser pulse which induces a vertical excitation of the neutral photoisomer $A$, Fig.~\ref{fig1}(c). The initial density matrix is therefore
\begin{equation}
    \rho(0) = \lvert \mathrm{S_B}\rangle \lvert \textbf{0}\rangle \langle \textbf{0}\rvert \langle \mathrm{S_B}\rvert~,
\end{equation}
where $ \lvert \textbf{0}\rangle$ denotes the vibrational ground state of the diabatic state $\lvert \mathrm{S_A}\rangle$.

\section{Results and discussion}
\subsection{Model system}
\begin{figure}
    \centering
    \includegraphics{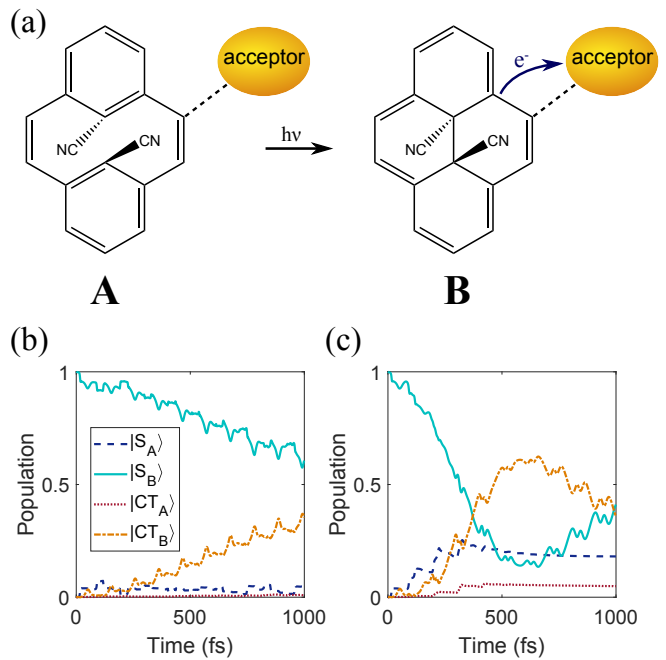}
    \caption{(a) Schematic of the studied model system. (b, c) Populations of the diabatic states as a function of time in the (b) absence, and (c) presence of vibrational damping $\gamma_k = 2$ meV and $T=300$ K.}
    \label{fig2}
\end{figure}

We demonstrate the principle of photoisomerization-coupled electron transfer on a concrete example although most of our conclusions are general.
We consider a system akin to that studied by Liddell \textit{et al.}\cite{liddell2004photonic} which comprises a `conventional' acceptor unit \footnote{In the experiments of Liddell \textit{et al.}, the acceptor unit was a cationic porphyrin unit which was prepared by a photoexciation of the porphyrin followed by an electron transfer to a fullerene unit.} and a photochromic dihydropyrene-type molecule acting as a (potential) electron donor.\cite{ayub2008suppressing}
As shown in Fig.~\ref{fig2}(a), the two possible photoisomers of the donor unit (\textbf{A} and \textbf{B}) possess different oxidation potentials so that photoisomeristion of the donor can be used to control the electron transfer.
We note that the considered photochromic unit belongs to the $C_{2h}$ point group in both of its neutral isomeric forms. We do not consider any linking groups in our calculations in order to make full use of the molecular symmetry.

We parameterise our (minimal and prototypical) model as follows.
First, geometry relaxation of the neutral and cationic forms of the two photoisomers is performed to obtain the energy differences between the two isomers in their respective charge states (all calculations are performed in NWChem at the B3LYP/6-311+G** level of theory\cite{valiev2010nwchem}). The relative alignment of the neutral and charge transfer states is chosen such that the ground states of $\lvert \mathrm{CT_A}\rangle$ and $\lvert \mathrm{CT_B}\rangle$ lie above and below those of the $\lvert \mathrm{S_A}\rangle$ and $\lvert \mathrm{S_B}\rangle$ states, respectively: $\bar{\varepsilon}_{\mathrm{S_A}} = 0$, $\bar{\varepsilon}_\mathrm{S_B} = -0.42$, $\bar{\varepsilon}_\mathrm{CT_A} = 0.32$, and $\bar{\varepsilon}_\mathrm{CT_B} = -0.5$ (all in eV).

Symmetry analysis reveals that the vibrational modes are $\Gamma_\mathrm{vib} = 24 A_g + 18 B_g + 19 A_u + 23 B_u$. The totally-symmetric $A_g$ modes can be easily identified as the tuning modes,\cite{kouppel1984multimode} whereas, as discussed in the supplementary material, the coupling modes are most likely of $B_u$ symmetry.
Since the electronic ground states of \textbf{A}, \textbf{B} and \textbf{B}$^\textbf{+}$ are all of $A_g$ symmetry, we indeed find that only the $A_g$ modes are displaced between the \textbf{A} and \textbf{B}/\textbf{B}$^\textbf{+}$ isomers. Meanwhile, the \textbf{A}$^\textbf{+}$ structure distorts into the $C_2$ symmetry which is accompanied by a displacement of the $A_g$ and $A_u$ modes. Consequently, no displacement of the coupling mode is associated with the electron transfer from the photochromic unit, $\kappa^{(j)}_\mathrm{c} =0$.
In order to determine the displacement parameters $\kappa^{(i)}_\mathrm{t}$, we use the neutral \textbf{A} structure as the reference geometry and follow a Dushinsky-type procedure as outlined in the supplementary material.
We choose the $A_g$ mode with the largest displacement for the $\textbf{A}\rightarrow \textbf{B}$ transition to act as the tuning mode in our calculation. This mode corresponds to the stretching/compressing of the carbon-carbon bond formed during the photoisomerization (see supplementary material) and has a frequency of $\omega_\mathrm{t} = 39.5$ meV. The relative displacements are $\kappa_\mathrm{t}^{(\mathrm{S_A})} = -0.235$, $\kappa_\mathrm{t}^{(\mathrm{S_B})} = 0.235$, $\kappa_\mathrm{t}^{(\mathrm{CT_A})} = -0.217$, and $\kappa_\mathrm{t}^{(\mathrm{CT_B})} = 0.217$ (all in eV).  
The resulting alignment of the energy levels is very similar to that shown in Fig.~\ref{fig1}(c), see supplementary material.
Crucially, therefore, the charge transfer is accompanied by a displacement of the tuning (isomerization) coordinate.

We do not attempt here to extract the interstate couplings $\lambda_i$ (or $\omega_\mathrm{c}$) from \textit{ab initio} calculations, \textit{cf.}~Ref.~\onlinecite{lan2008photochemistry}, resorting instead to values previously suggested in the literature,\cite{duan2018signature} $\omega_\mathrm{c} = 0.112$ eV and $\lambda_0 = \lambda_1 = 49.6$ meV, assuming for simplicity that inter-state coupling is identical in the neutral and charge-transfer states.
To further reduce the number of parameters, we set $J_{ij} = J = 5$ meV, unless stated otherwise.

\subsection{Properties of the model}
Let us begin by exploring the properties of our model system.
(i) Firstly, in the absence of the initial photoexcitation, that is, when $\rho(0) = \lvert \mathrm{S_A}\rangle \lvert \textbf{0}\rangle \langle \textbf{0}\rvert \langle \mathrm{S_A}\rvert$, the electron transfer is not energetically possible as the $\lvert \mathrm{CT}_\mathrm{A}\rangle$ state lies above the $\lvert \mathrm{S}_\mathrm{A}\rangle$ state and there exists an energy barrier for the direct isomerization. Consequently, on the considered timescale, the system essentially remains in the (meta-stable) $\lvert \mathrm{S_A}\rangle$ state, see supplementary material.
(ii) In the absence of donor-acceptor electronic coupling, that is for $J=0$, following a photoexciation of the \textbf{A} isomer the molecular system simply isomerises from \textbf{A} to \textbf{B} with a yield of roughly $0.74$ (for $\gamma_k = 2$ meV and temperature $T=300$ K). As explicitly shown in the supplementary material, this process takes place within less than 1 ps.
(iii) For the system initially found in the vibrational ground state of state $\lvert \mathrm{S_B}\rangle$, the electron transfer to the state $\lvert \mathrm{CT_B}\rangle$ takes place without the need for laser excitation due to the relatively small energy barrier between the $\lvert \mathrm{S_B}\rangle$ and $\lvert \mathrm{CT_B}\rangle$ states, see Fig.~\ref{fig1}(c) and the supplementary material.
(iv) Finally, as we shall discuss in the remainder of this work, the photoexcitation of the wavepacket located in the ground state of the $\lvert \mathrm{S_A}\rangle$ state triggers a simultaneous isomerization of the photochromic donor moiety and an electron transfer from that donor to the acceptor unit.

\subsection{Dynamics}
\begin{figure}[t]
    \centering
    \includegraphics{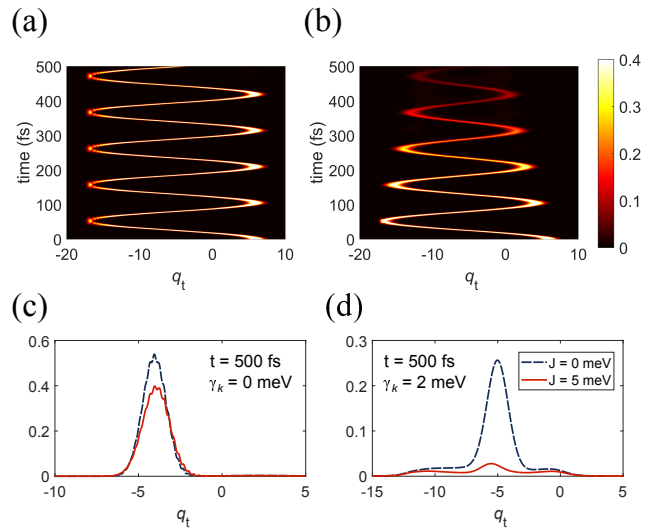}
    \caption{(a, b) Projections of the wavepacket on the tuning coordinate as a function of time for (a) $\gamma_k = 0$, and (b) $\gamma_k = 2$ meV. (c, d) Projections of the wavepacket on the tuning coordinate after 500 fs of wavepacket evolution for (c) $\gamma_k = 0$, and (d) $\gamma_k = 2$ meV; $T = 300$ K. }
    \label{fig3}
\end{figure}
We first consider the fully unitary dynamics, that is, we set $\gamma_k = 0$. As shown in Fig.~\ref{fig2}(b), following the vertical photoexcitation, we observe coherent oscillations of the populations of the four diabatic states in which the small initial rise of the population of the $\lvert \mathrm{S_A}\rangle$ state is followed by a fast increase of the charge-transfer states.
\begin{figure*}
    \centering
    \includegraphics{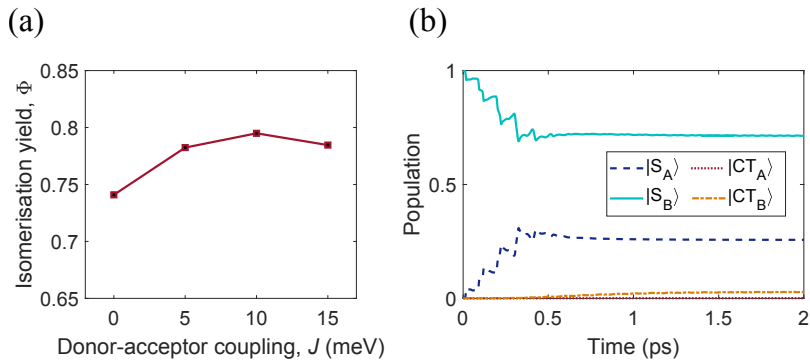}
    \caption{(a) Isomerization yield $\Phi$ as a function of the donor-acceptor coupling, $J$. (b) Diabatic population dynamics for very weak donor-acceptor coupling, $J = 0.4$ meV. Remaining parameters as in Fig.~\ref{fig2}(c).}
    \label{fig4}
\end{figure*}
As shown in Fig.~\ref{fig2}(c), in the presence of vibrational dissipation the initial coherent oscillations are largely damped which is accompanied by a more efficient electron transfer.
The photoisomerization takes place  within roughly 600 fs. This is accompanied by the levelling off of the populations of the $\lvert \mathrm{S_A}\rangle$ state (formed by a propagation of the vibrational wavepacket through the conical intersection).
In the case of a successful photoisomerization, the `excess' electron is at longer times delocalised over  the non-photochromic acceptor and the donor unit (in its $B$ form), see the supplementary material for the population dynamics at longer times. 
This occurs due to relatively strong donor-acceptor coupling that is comparable to the energy difference between the $\lvert \mathrm{S_B}\rangle$ and $\lvert \mathrm{CT_B}\rangle$ states in our model system (and is not therefore universally true).
Surprisingly, we also observe a small (but non-negligible) population of the $\lvert \mathrm{CT_A}\rangle$ state. This corresponds to the electron transfer having taken place despite an unsuccessful photoisomerization of the donor unit. Generally, this has a chance of occurring as long as the photoinduced wavepacket propagates through the (avoided) crossing of the $\lvert \mathrm{S_B}\rangle$ and $\lvert \mathrm{CT_A}\rangle$ diabatic states, as it can be inferred from Fig.~\ref{fig1}(c).

We next demonstrate the inseparability of the photoisomerization and electron transfer processes.
First, in Figs.~\ref{fig3}(a, b), we plot the projection of the wavepacket in the diabatic state $\lvert \mathrm{S_B}\rangle$ on the tuning coordinate as a function of time. We observe the expected oscillations of the wavepacket, damped towards $q_\mathrm{t} = \Delta_\mathrm{S_B}$ in the case of non-zero $\gamma_\mathrm{t}$. Figs.~\ref{fig3}(c) and (d) show the projection of the wavepacket at time $t=500$ fs for two values of the donor-acceptor coupling: $J=0$ and 5 meV. Crucially, the possibility of the electron transfer (non-zero $J$) affects the shape as well as the magnitude of the wavepacket. The shape of the wavepacket (after certain propagation time) depends non-trivially on the details of the potential energy surface.  The decrease in its magnitude, on the other hand, can be easily understood as a consequence of populating the charge transfer states.

It is also useful to consider the yield of photoisomerization which one can define (only in the presence of vibrational damping) as $\Phi = P_\mathrm{S_B}(t_{ss}) + P_\mathrm{CT_B}(t_{ss})$ where $P_i(t) = \mathrm{Tr}\left[\lvert i\rangle \langle i\rvert \rho(t)\right]$ and $t_{ss}$ is the time at which the dynamics of the system reaches its quasi-steady-state.\cite{schile2019simulating} For the parameters used here $t_{ss} \sim 3$ ps (see supplementary material for the population dynamics for the considered values of the donor-acceptor coupling). Fig.~\ref{fig4}(a) shows the photoisomerization yield as a function of the donor-acceptor coupling, $J$.
The initial increase of $\Phi$ with increasing $J$ is due to the possibility of charge transfer which leads to populating the (energetically more stable) $\lvert\mathrm{CT_B}\rangle$ state.
On the other hand, for larger values of the donor-acceptor coupling, the photoisomerization yield marginally decreases which can be attributed to the formation of the $\lvert \mathrm{CT_A}\rangle$ products.
Crucially, however, Fig.~\ref{fig4}(a) reveals that the photoisomerization yield depends on the strength of the donor-acceptor coupling. For the parameters (and the model system) considered here, the changes of $\Phi$ with $J$ are relatively modest. It should nevertheless be possible to detect them experimentally, see Ref.~\onlinecite{ladanyi2017azobenzene}.
Since the possibility of charge transfer from the photochromic donor unit affects the overall photoisomerization yield, the two processes clearly need to be treated within a unified theory, as asserted above.

Finally, we turn to the issue of controlling the efficiency of the photoisomerization-coupled electron transfer. As expected, the rate of the considered electron transfer decreases with decreasing donor-acceptor coupling $J$.
As discussed above and  shown in Fig.~\ref{fig4}(b), in the case of significantly smaller donor-acceptor coupling, the photoisomerization and electron transfer take place on very different timescales. The electron transfer (which can be observed as the rise of the populations of the charge transfer states) takes place largely after the photoisomerization is complete (that is, when the $\lvert \mathrm{S_A}\rangle$ and $\lvert \mathrm{S_B}\rangle$ populations have almost reached their quasi-steady-state values). Consequently, the electron transfer and photoisomerization may essentially be treated as separate processes.
It is also worth noting that in the case of much weaker donor-acceptor coupling, no significant populations of the $\lvert \mathrm{CT_A}\rangle$ state are observed, in contrast to what was discussed above. 
It is important to note here that the donor-acceptor coupling can be tuned (or at least varied) experimentally by changing the chemical group linking the donor and acceptor units.

Although the wavepacket dynamics are sensitive to the details of the potential energy surfaces the latter are inherent to the considered photochromic system. Similarly, environmental interactions strongly influence the dynamics of the photoisomerization-coupled electron transfer, but are generally difficult to control experimentally. Nonetheless, the efficiency of the electron transfer can be to some extent controlled by changing the offset between the neutral and charge-transfer states. In practice this can be achieved by varying the non-photochromic acceptor (or, if appropriate, donor) moiety.
As can be inferred from Fig.~\ref{fig1}(c), by changing the relative energies of the neutral and the charge transfer states, one can modify the energy barrier for the electron transfer for the isomer B.
In particular, as we demonstrate in the supplementary material, stabilising the charge transfer states (that is, lowering their energy relative to that of neutral singlet states) can push the $\lvert \mathrm{S_B}\rangle \rightarrow \lvert \mathrm{CT_B}\rangle$  electron transfer into a deep inverted region thus impairing the efficiency of this process.
In the context of controlling the dynamics of photoisomerization-coupled electron transfer, the use of colloidal quantum dots as the non-photochromic acceptor (or donor) units\cite{erno2010optical} is therefore particularly attractive due to the apparent tunability of these systems. (It is possible to shift the positions of the quantum-dot  electronic energy levels  by simply changing the size of these particles.)

\section{Conclusions}
In this work, we have introduced and developed a conceptual understanding of the phenomenon of photoisomerization-coupled electron transfer which amounts to a simultaneous and inseparable photoisomersation and electron transfer from (or onto) a photochromic donor (or acceptor) unit. 
We have demonstrated, using the example of a dihydropyrene-type molecular structure, how this process can be modelled and, to some extent, controlled by the chemical design of the donor-acceptor system. 
We have also shown that this phenomenon can yield products inaccessible through a sequential isomerization and charge transfer reactions.
Furthermore, we discussed how the photoisomerization yield can be affected by the possibility of an electron transfer. If a stronger dependence of the photoisomerization yield on the donor-acceptor coupling (than observed in our prototypical model system) can be realised, the process considered here may open an attractive new way of controlling the outcomes of photoisomerization reactions. 
An understanding of the dynamics of photoisomerization-coupled electron transfer should also prove crucial for the future design of molecular switches and memory devices based on the phenomenon considered here, especially when an ultrafast operation of such devices is desirable.

While awaiting experimental studies of the considered phenomenon, several aspects of photoisomerization-coupled electron transfer can be further explored theoretically. These include extending the theory presented here to the multi-mode case, anharmonic potential energy surfaces and \textit{cis-trans} isomerization as well as identifying molecular structures suitable for experimental investigations.
Designing systems in which  photoisomerization and electron transfer are maximally intertwined would be especially important for the future empirical research into the process considered here.
It would likewise be interesting to investigate in more detail the role of environmental interactions (also beyond the weak-coupling approximation) including the importance of possible non-Markovian effects.
Finally, \textit{ab initio} molecular dynamics simulations could also be used to study the phenomenon in question.

\section*{Supplementary material}
See supplementary material for several further examples of photochromic units, details of the theoretical methods, and supporting dynamics calculations.

\begin{acknowledgements}
The authors thank Roel Tempelaar for comments on the manuscript.
This work was supported by the National Science Foundation’s MRSEC program (DMR-1720139) at the Materials Research Center of Northwestern University.
This research was also supported in part through the computational resources and staff contributions provided for the Quest high performance computing facility at Northwestern University which is jointly supported by the Office of the Provost, the Office for Research, and Northwestern University Information Technology.\\
\end{acknowledgements}

\section*{Data availability statement}
The data that support the findings of this study are available within the article and its supplementary material.

\end{document}